\documentclass[%
reprint, %linenumbers,
superscriptaddress,
footinbib,
amsmath,amssymb,
aps, prd,
]{revtex4-2}
\usepackage{aas_macros}
\usepackage{graphicx}% Include figure files
\usepackage{dcolumn}% Align table columns on decimal point
\usepackage{bm}% bold math
\usepackage[colorlinks=true, allcolors=blue]{hyperref}
\begin{document}

\title{Bright flare in the obscured state of GRS 1915+105 as seen by NICER and Swift}

\author{Shuaitongze~Zhao}
\affiliation{Center for Astronomy and Astrophysics, Department of Physics, Fudan University, Shanghai 200438, China}
\affiliation{Institut f\"ur Astronomie und Astrophysik, Eberhard-Karls Universit\"at T\"ubingen, D-72076 T\"ubingen, Germany}

\author{Honghui~Liu}
\email[Corresponding author: ]{honghui.liu@uni-tuebingen.de}
\affiliation{Institut f\"ur Astronomie und Astrophysik, Eberhard-Karls Universit\"at T\"ubingen, D-72076 T\"ubingen, Germany}

\author{Menglei~Zhou}
\affiliation{Institut f\"ur Astronomie und Astrophysik, Eberhard-Karls Universit\"at T\"ubingen, D-72076 T\"ubingen, Germany}

\author{Swarnim~Shashank}
\affiliation{Institut f\"ur Astronomie und Astrophysik, Eberhard-Karls Universit\"at T\"ubingen, D-72076 T\"ubingen, Germany}
\affiliation{College of Fellows, Eberhard-Karls Universit\"at Tübingen, D-72070 T\"ubingen, Germany}
\affiliation{Center for Astronomy and Astrophysics, Department of Physics, Fudan University, Shanghai 200438, China}

\author{Cosimo~Bambi}
\email[Corresponding author: ]{bambi@fudan.edu.cn}
\affiliation{Center for Astronomy and Astrophysics, Department of Physics, Fudan University, Shanghai 200438, China}
\affiliation{School of Humanities and Natural Sciences, New Uzbekistan University, Tashkent 100000, Uzbekistan}

\author{Andrea~Santangelo}
\affiliation{Institut f\"ur Astronomie und Astrophysik, Eberhard-Karls Universit\"at T\"ubingen, D-72076 T\"ubingen, Germany}
\affiliation{Center for Astronomy and Astrophysics, Department of Physics, Fudan University, Shanghai 200438, China}

\begin{abstract}
We report time-resolved NICER and Swift X-ray spectroscopy of a bright flare from the black hole X-ray binary GRS~1915+105 during its obscured state, which is characterized by heavy line-of-sight absorption by dense material with complex geometry. In April 2023, an unexpected flare was detected, with the observed X-ray flux increasing by nearly an order of magnitude relative to the typical obscured-state level. The spectra show pronounced variability, including significant evolution of the Fe~K emission features. Time-resolved spectral modeling indicates that the main flare is associated with a combination of enhanced intrinsic emission and reduced obscuration. We further find that neutral and ionized reflection components are subject to distinct absorbers, whose evolving visibility implies a stratified absorber–reflector geometry. These properties are consistent with a re-illumination phase following a failed disk wind. A delayed radio flare detected about 2.5 days later suggests a coupling between accretion and jet activity. 
\end{abstract}

\maketitle

%%%%%%%%%%%%%%%%%%%%%%%%%%%%%%%

\section{Introduction} \label{sec:style}

GRS~1915+105 is one of the most luminous and variable black hole X-ray binaries (BHXRBs) in the Galaxy. Since its discovery in 1992 \cite{Castro-Tirado_1992, Castro-Tirado_1994}, the source has remained active for over 26 years. It has exhibited rich temporal and spectral behavior, including superluminal radio jets \cite{Mirabel_1994}, ionized accretion disk winds (e.g., \cite{Ueda_2009,neilsen_2009}), and rapid transitions between multiple variability classes \cite{Belloni_2000}.

In early 2018, the X-ray flux of GRS~1915+105 decreased unexpectedly \cite{Homan2019}. In April 2019, the source entered a previously unobserved low-flux state, referred to as the ``obscured state'' \cite{Miller_2020}. Unlike the canonical low luminosity hard state observed in BHXRBs, this state of GRS1915+105 is characterized by an unusually high X-ray absorption column and Compton-thick obscuration (e.g., \cite{Miller_2020,Balakrishnan_2021,motta2021}). At the same time, mid-infrared and radio observations (e.g., \cite{motta2019,Trushkin_2023_3,Gandhi2025}) indicated ongoing accretion activity, suggesting that the observed low X-ray luminosity is not due to a low accretion rate but instead to heavy obscuration along the line of sight. Several physical scenarios have been proposed to explain the origin of the obscuration, including a failed disk wind \cite{Miller_2020}, vertical expansion of the outer accretion disk \cite{Neilsen2020}, and the accumulation of bound outflows from previous super-Eddington episodes \cite{Koljonen2020}. In 2023, mid-infrared observations with the James Webb Space Telescope (JWST) further suggested that the deep X-ray obscured state may be explained by alignment of the accretion disk plane with the line of sight \cite{rodriguez2025}. Radio observations also revealed a large misalignment between the continuous jet and discrete ejecta, suggesting a time-variable jet-launching geometry that may be connected to a warped or distorted accretion disk \citep{jiang2026}.

\begin{figure*}
    \centering
     \includegraphics[width=0.9\linewidth]{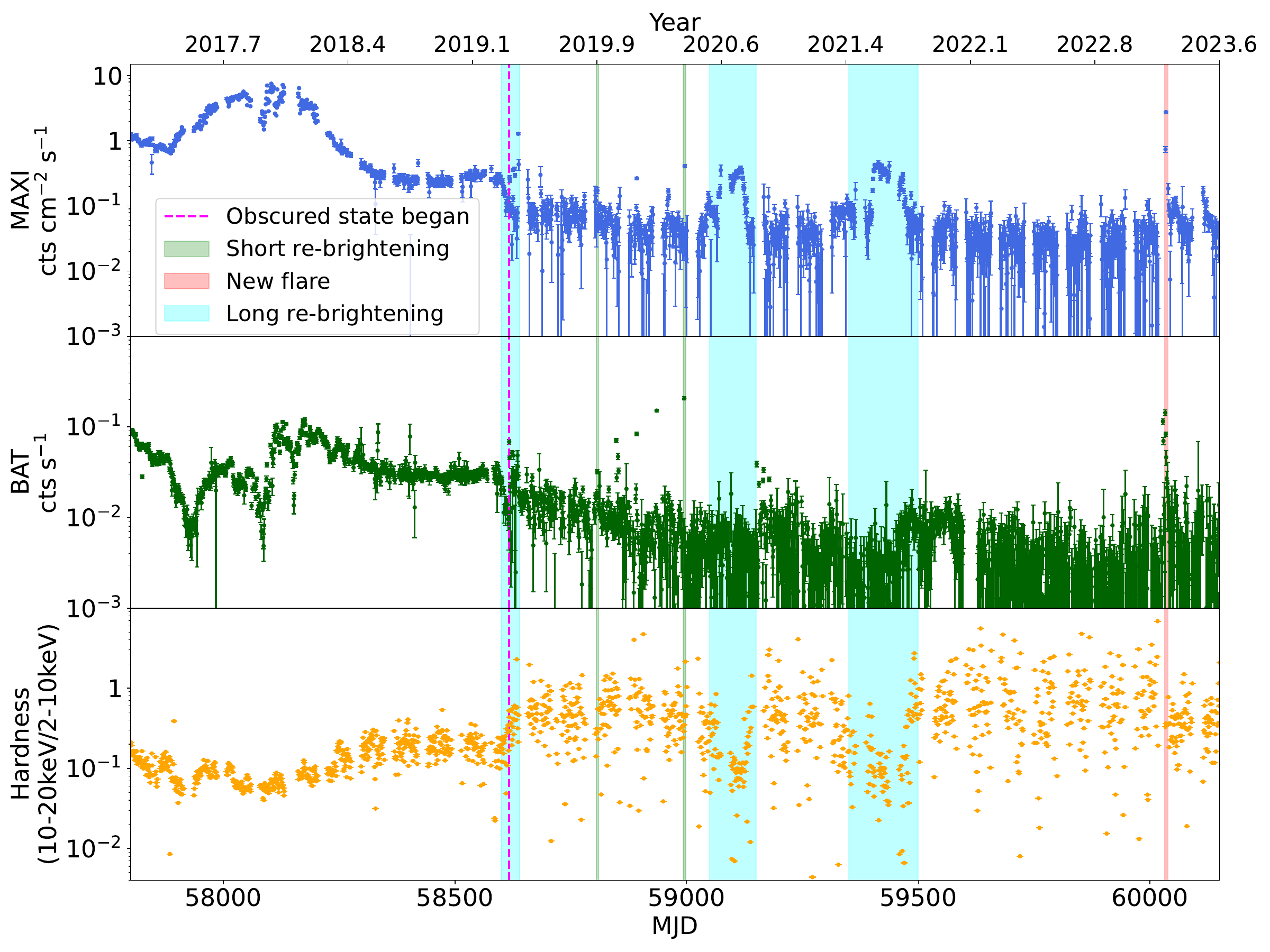}
    \caption{Top panel: MAXI 2--20\,keV flux of the source between MJD~57800 and MJD~60150. Central panel: BAT 15--150\,keV flux. Bottom panel: MAXI hardness ratio (10--20\,keV / 2--10\,keV). The onset of the obscured state is marked by the magenta dashed line. The cyan and green shaded regions indicate the long and short re-brightening events reported by \citet{Athulya2023}, respectively. The pink shaded region marks the new flare analyzed in this work observed by NICER and Swfit on MJD~60034.}
    \label{fig:maxi}
\end{figure*}

On March 29, 2023 (MJD 60034), an unexpected strong X-ray flare was detected \cite{Yamada2023} by both the Burst Alert Telescope (BAT; \cite{Barthelmy_2005_BAT}) onboard Swift and the Monitor of All-sky X-ray Image (MAXI; \cite{Matsuoka_2009_MAXI}) (Figure~\ref{fig:maxi}). The X-ray flux increased by more than an order of magnitude relative to the average level in the obscured state, with both the soft and hard bands becoming significantly brighter. Remarkably, the source flux temporarily returned to its pre-obscuration level during the flare. Also, the X-ray flare was followed by a large radio flare with a delay of about 2.5 days \cite{Trushkin_2023_2}, suggesting a potential link between the inner accretion flow and jet activity in the obscured state.

During one bright individual flare, how the absorber, continuum, and reflection components evolve remains less clear. In this work, we analyze the X-ray spectra of this bright, short-duration flare observed during the obscured state of GRS~1915+105. We find that the flare is driven by both intrinsic luminosity variations and changes in the local absorbers, with their relative importance evolving over time. The rise of the flare is primarily dominated by an increase in the intrinsic source luminosity, while the subsequent decay is governed by enhanced obscuration from the absorbers.
The paper is organized as follows. Section~\ref{sec:obs} describes the NICER and Swift/XRT data reduction procedures. In Section~\ref{sec:temp}, we present the evolution of the light curves and spectra. Our spectral modeling results are presented in Sections~\ref{sec:gau} and \ref{sec:ref}, and are discussed in Section~\ref{sec:dis}. The summary and conclusions are given in Section~\ref{sec:conclusion}.

\label{fig:1}

%%%%%%%%%%%%%%%%%%%%%%%%%%%%%%%

\section{Observations and Data Reduction}\label{sec:obs}
\subsection{NICER}
NICER \cite{Gendreau_2016} observed GRS~1915+105 during its obscured state, covering the energy range from 0.2 to 12\,keV. We focus on the flare observed between 2023 March 31 and April 1 (UTC), corresponding to ObsIDs 6103010201, 6103010202, and 6697010201, which span the rise, peak, and decay phases of the event.

The data were downloaded from the High Energy Astrophysics Science Archive Research Center (\texttt{HEASARC}). We processed the observations using \texttt{HEASoft} v6.32.1 and the NICER Calibration Database (\texttt{CALDB}) version \texttt{xti20240206}. Cleaned event files were generated with \texttt{nicerl2}, and detectors 14, 34, and 54 were excluded to reduce instrumental noise \cite{2019ApJ...887L..25B}. Light curves and spectra were extracted using \texttt{nicerl3-lc} and \texttt{nicerl3-spect}, respectively.

Spectra were divided according to NICER's orbital visibility, with segments separated when the time gap between Good Time Intervals (GTIs) exceeded 1000\,s. In addition, spectra with a response below 10\% of the on-axis value were discarded to ensure reliable calibration. For all intervals, we applied the optimal binning algorithm \cite{Kaastra2016}. The spectral analysis is performed in the 1--10\,keV energy range. 

\subsection{Neil Gehrels Swift Observatory}

Swift/XRT \cite{Burrows_2005} monitored GRS~1915+105 on 2023 March 31 (ObsID 00012178134), following the main flare observed by NICER. The data are free of pile-up effects. Cleaned event files were generated using \texttt{xrtpipeline} v0.13.7. Source spectra were extracted from a circular region with a radius of 30$^{\prime\prime}$, and background spectra were extracted from a nearby source-free region with a radius of 35$^{\prime\prime}$. Ancillary response files (ARFs) were generated using \texttt{xrtmkarf}.

\begin{figure*}
    \centering
     \includegraphics[width=0.7\linewidth]{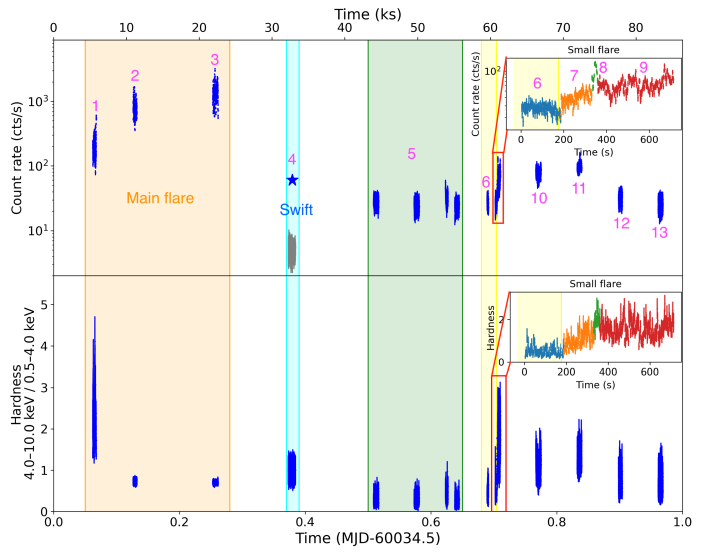}
    \caption{Upper panel: Light curves of the observations, divided into 16 intervals based on time and hardness. The main flare occurs in intervals 1–3, while a smaller flare is observed in intervals 6–9 (shown in cyan, orange, green, and red, respectively). Swift counts are shown in grey. The roughly rescaled Swift point, estimated from the flux, is shown as a blue star. Bottom panel: Hardness ratio (4--10\,keV / 0.5--4\,keV) of the observations.}
    \label{fig:all_lc}
\end{figure*}

%%%%%%%%%%%%%%%%%%%%%%%%%%%%%%%

\section{Time-Resolved Spectroscopy} \label{sec:temp}

\subsection{Light curve and spectra}

The light curves are divided into thirteen intervals based on NICER's orbital visibility, as shown in Fig.~\ref{fig:all_lc}. Among these, intervals 1–3 correspond to the main flare, during which the X-ray flux increases by approximately one order of magnitude. Intervals 6–9, which correspond to a smaller flare, are further subdivided according to noticeable changes in the hardness ratio. The four exposures in interval 5 and the two exposures in interval 6 are combined to improve the signal-to-noise ratio after verifying the spectral variability. At the onset of the small flare, the hardness is low and nearly constant (cyan). As the count rate increases, the hardness rises (orange). When the flare reaches its peak, the hardness becomes even higher (green). After the small flare, both the light curve and the hardness remain variable (red).

\begin{figure}
    \centering
    \includegraphics[width=1\linewidth]{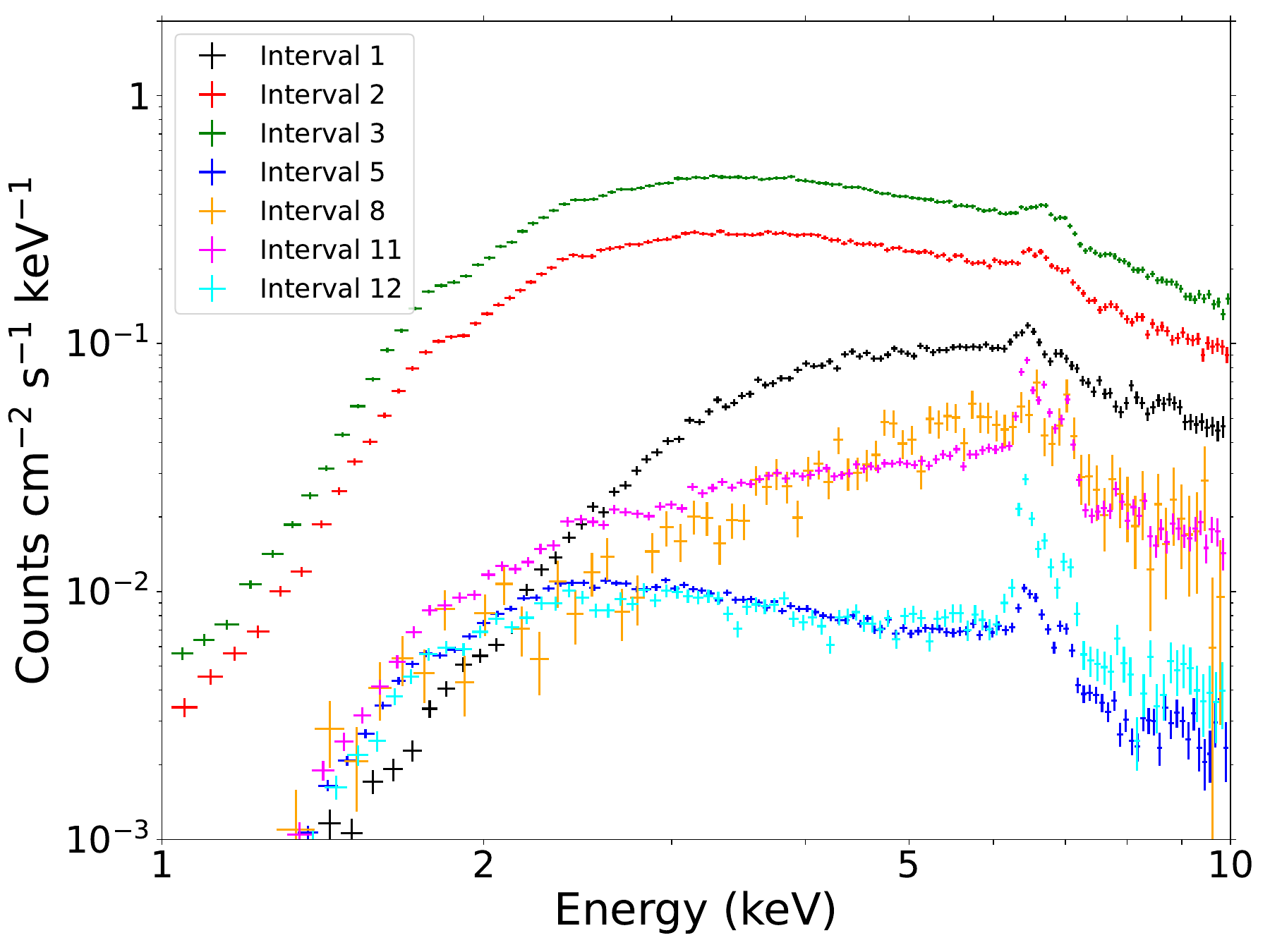}
    \caption{Typical spectra for our observations. Different colors stand for different intervals (Black, red, green, blue, orange, magenta, cyan for interval 1, 2, 3, 5, 8, 11, 12, respectively). The sequence is according to the number shown in Figure \ref{fig:all_lc}. }

    \label{fig:eeuf}
\end{figure}

The flares exhibit strong variability, as seen in both the light curves (Figure~\ref{fig:all_lc}) and the spectra (Figure~\ref{fig:eeuf}). Interval~1 marks the onset of the main flare, which is followed by a steady softening of the spectrum as the count rate increases (intervals 2 and 3). The flare peaks in interval~3, reaching nearly 2000~cts\,s$^{-1}$ for NICER, corresponding to approximately ten times the flux in interval~1. The flux then drops sharply to about 30~cts\,s$^{-1}$ in interval~5, during which the hardness remains roughly constant, albeit with larger uncertainties. A smaller flare is observed between intervals~6 and~8, where the count rate increases from $\sim$30 to $\sim$140~cts\,s$^{-1}$ within $\sim$300~s, accompanied by a hardening of the spectrum. In interval~9, the flux declines slightly without a corresponding change in the hardness. Interestingly, from intervals~4 to~9, the spectral shape between 1 and 3~keV remains relatively unchanged, whereas the flux above 4~keV shows significant variability. Similar behavior is also observed between intervals~10 and~11, as well as between intervals~12 and~13. This phenomenon has been reported in previous studies (e.g., \cite{Neilsen2020}). From intervals~10 to~13, both the flux and hardness continue to decrease.

\subsection{Spectral analysis and modeling}

We performed all spectral analyses with \textsc{xspec} v12.13.1 and $\chi^2$ statistics.

Interstellar photoelectric absorption was modeled with \texttt{TBabs}, adopting the abundances of \citet{Wilms2000}. We initially attempted to fit the spectra with a multicolor disk blackbody model only (\texttt{diskbb}; \cite{Mitsuda1984, Makishima1986}), but failed to obtain good fits. 

We then applied a thermal Comptonization model (\texttt{nthComp}; \cite{Zdziarski1996, zycki1999}), as well as a combined \texttt{diskbb + nthComp} model, but neither provided a statistically significant improvement to the fits. Additionally, we tested the \texttt{simplcutx} $\times$ \texttt{diskbb} model \cite{Steiner2017}, but it failed to yield physically reasonable results in the first three intervals, producing unrealistically low disk temperatures and excessively large $R_{\mathrm{in}}$ values. These results indicate that the Comptonized continuum emission with a simple neutral absorber alone cannot adequately describe the spectra, motivating the inclusion of an additional partial-covering absorber \cite[e.g.,][]{motta2021}.

\begin{figure*}
    \centering
     \includegraphics[width=1\linewidth]{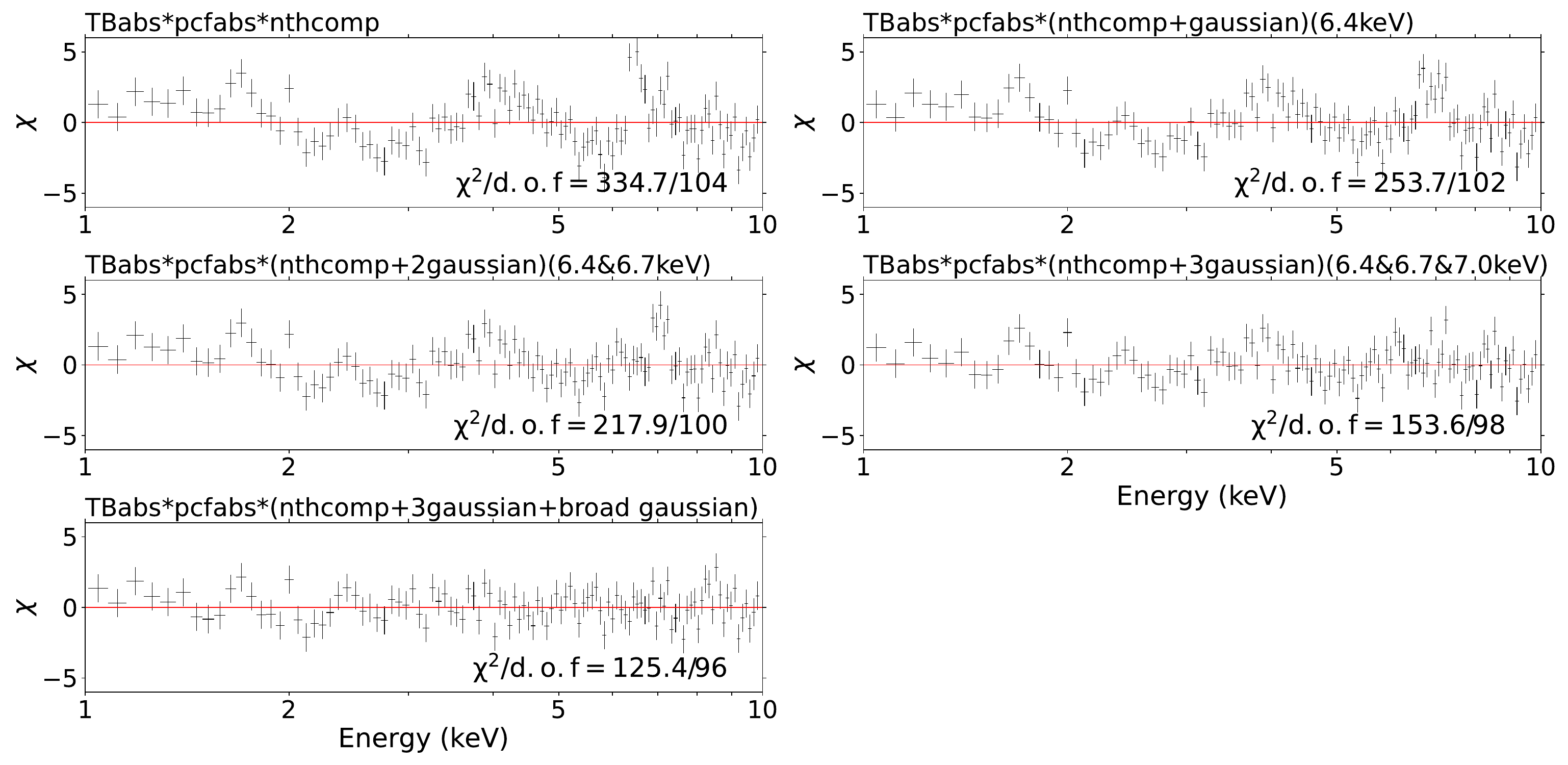}
     \caption{Examples of spectral residuals for interval~9. The residuals are shown in black. The upper-left panel shows the residual when fitting data with a baseline model, \texttt{TBabs*pcfabs*nthcomp}, which reveals prominent residuals at 6.4\, 6.7\, and 7.0\,keV. The line energies were allowed to vary within $\pm$0.05\,keV. For each narrow line, the line width was fixed at $\sigma = 0.01$\,keV. The broad \texttt{Gaussian} component was fixed at 6.4\,keV with $\sigma > 0.1$\,keV.  }
    \label{fig:12eemo}
\end{figure*}

We modeled the spectra using a continuum model of \texttt{TBabs} $\times$ \texttt{pcfabs} $\times$ (\texttt{nthComp} $+$ Gaussian components), hereafter referred to as the Gaussian-based model. The \texttt{TBabs} component was linked across all 13 intervals, and the temperature of the seed photons emitted from the disk was fixed at 0.1\,keV. After fitting the continuum, we inspected the spectral residuals, which reveal three distinct features near 6.4, 6.7 and 7.0~keV. These correspond to neutral Fe~K$\alpha$, He-like Fe~\textsc{xxv}, and either H-like Fe~\textsc{xxvi} Ly$\alpha$ or Fe~K$\beta$ emission lines, respectively, as illustrated for interval~9 in Figure~\ref{fig:12eemo}. Each feature was modeled with a narrow Gaussian line with a fixed width of $\sigma = 0.01$~keV, while the line energies were allowed to vary within $\pm$0.05~keV.
For some intervals, an additional broad iron line at 6.4~keV with $\sigma > 0.1$~keV was required. Gaussian line components were included only when their addition improved the fit at a significance level exceeding 3$\sigma$, based on the $\Delta\chi^2$ criterion. The significance is calculated based on the \texttt{error} command. Consequently, the number of required lines varies between intervals. 

\subsection{Fitting results}\label{sec:gau}
\begin{figure}
    \centering
    \includegraphics[width=1\linewidth]{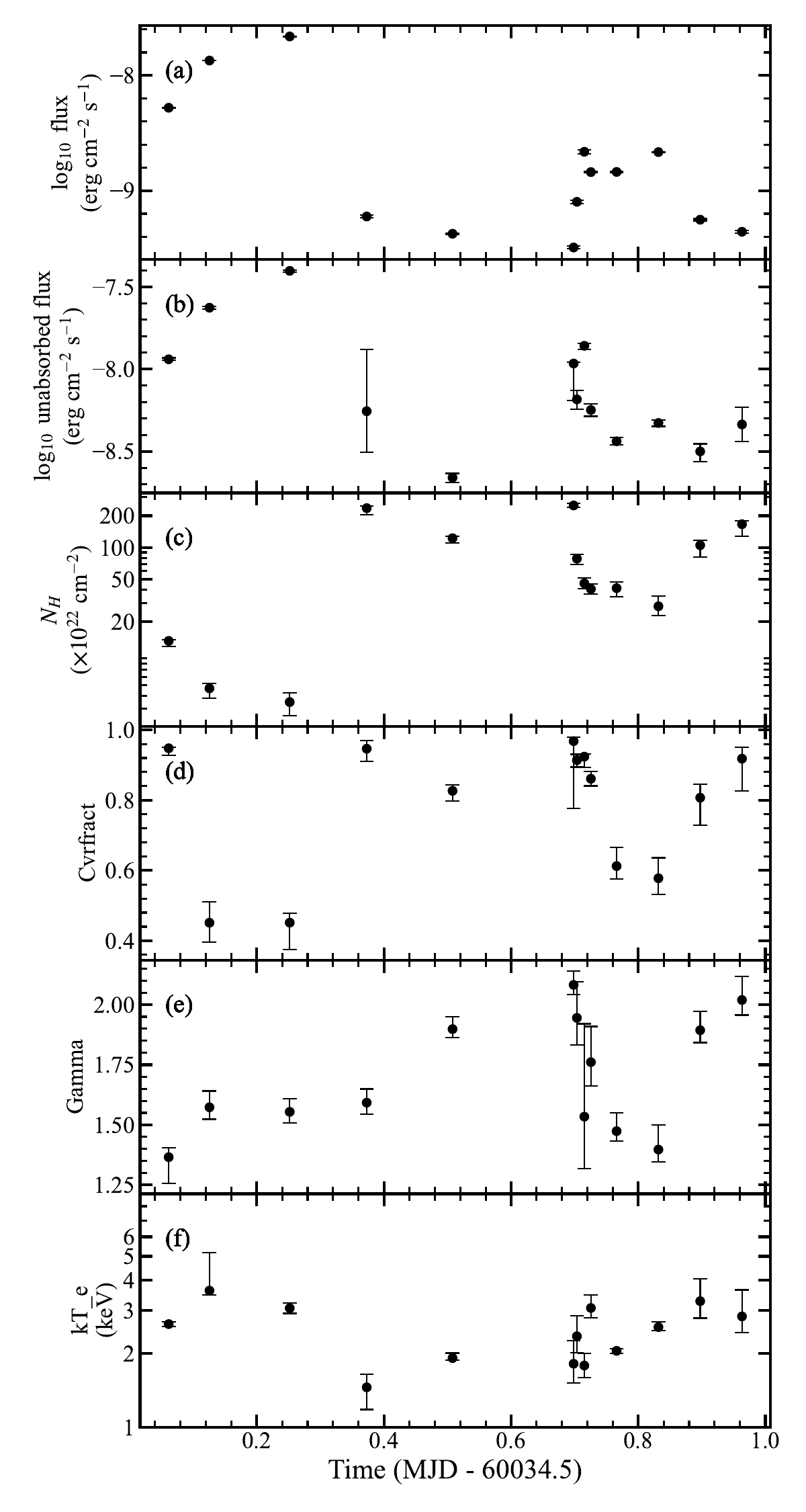}
    \caption{The evolution of the parameters from the fitting results of model \texttt{TBabs} $\times$ (\texttt{pcfabs} $\times$ \texttt{nthComp}$+$ \texttt{gaussians}). (a) observed flux from 1\,keV to 10\,keV (b) unabsorbed flux from 1\,keV to 10\,keV; (c) obscuring column density $N_{\mathrm H}$; (d) covering fraction $Cvrfract$; (f) photon index $\Gamma$; (g) electron temperature $\mathrm{kT_ e}$. 
    }
    \label{fig:par}
\end{figure}

\begin{table*}
\centering
\renewcommand\arraystretch{1.5}
\begin{tabular}{lcccccc}
\hline\hline  
\shortstack[l]{Interval \\~~}
&
\shortstack[l]{$\log_{10}F$\\($\mathrm{erg\,cm^{-2}\,s^{-1}}$)}
&
\shortstack[l]{$\log_{10}F_{\mathrm{unabs}}$\\($\mathrm{erg\,cm^{-2}\,s^{-1}}$)}
&
\shortstack[l]{$CvrFract$ \\~~}
&
\shortstack[l]{$N_{\mathrm H}$\\($10^{22}\,\mathrm{cm^{-2}}$)}
&
\shortstack[l]{$\Gamma$ \\~~}
&
\shortstack[l]{$\mathrm{kT_ e}$\\($\mathrm{keV}$)} \\
\hline

1 & $-8.280^{+0.003}_{-0.003}$ & $-7.940^{+0.009}_{-0.004}$ & $0.953^{+0.003}_{-0.02}$ & $13.1^{+0.4}_{-1.5}$ & $1.37^{+0.04}_{-0.1}$ & $2.6^{+0.06}_{-0.06}$ \\
2 & $-7.870^{+0.002}_{-0.002}$ & $-7.626^{+0.007}_{-0.008}$ & $0.45^{+0.06}_{-0.06}$ & $4.7^{+0.5}_{-0.9}$ & $1.57^{+0.07}_{-0.05}$ & $3.6^{+1.6}_{-0.1}$ \\
3 & $-7.661^{+0.001}_{-0.001}$ & $-7.402^{+0.006}_{-0.006}$ & $0.45^{+0.03}_{-0.08}$ & $3.4^{+0.8}_{-0.9}$ & $1.55^{+0.06}_{-0.04}$ & $3.1^{+0.1}_{-0.1}$ \\
4 & $-9.22^{+0.01}_{-0.01}$ & $-8.3^{+0.4}_{-0.2}$ & $0.95^{+0.02}_{-0.04}$ & $235^{+10}_{-31}$ & $1.59^{+0.06}_{-0.05}$ & $1.92^{+0.09}_{-0.04}$ \\
5 & $-9.374^{+0.004}_{-0.004}$ & $-8.65^{+0.02}_{-0.03}$ & $0.83^{+0.02}_{-0.03}$ & $122^{+5}_{-11}$ & $1.90^{+0.05}_{-0.04}$ & $1.5^{+0.2}_{-0.3}$ \\
6 & $-9.49^{+0.01}_{-0.01}$ & $-7.966^{+0.008}_{-0.2}$ & $0.97^{+0.01}_{-0.2}$ & $249^{+11}_{-10}$ & $2.08^{+0.06}_{-0.04}$ & $1.8^{+0.4}_{-0.3}$ \\
7 & $-9.10^{+0.01}_{-0.02}$ & $-8.18^{+0.06}_{-0.06}$ & $0.91^{+0.02}_{-0.02}$ & $78^{+7}_{-10}$ & $2.0^{+0.2}_{-0.1}$ & $2.4^{+0.5}_{-0.3}$ \\
8 & $-8.66^{+0.02}_{-0.02}$ & $-8.06^{+0.09}_{-0.09}$& $0.923^{+0.008}_{-0.03}$ & $46^{+6}_{-5}$ & $1.5^{+0.4}_{-0.2}$ & $1.8^{+0.2}_{-0.2}$ \\
9 & $-8.838^{+0.007}_{-0.007}$ & $-8.25^{+0.04}_{-0.04}$ & $0.86^{+0.02}_{-0.02}$ & $40^{+5}_{-4}$ & $1.8^{+0.1}_{-0.1}$ & $3.1^{+0.4}_{-0.3}$ \\
10 & $-8.837^{+0.004}_{-0.005}$ & $-8.44^{+0.02}_{-0.02}$ & $0.61^{+0.05}_{-0.04}$ & $41^{+6}_{-7}$ & $1.47^{+0.08}_{-0.04}$ & $2.05^{+0.04}_{-0.05}$ \\
11 & $-8.665^{+0.004}_{-0.004}$ & $-8.33^{+0.02}_{-0.02}$ & $0.58^{+0.06}_{-0.04}$ & $28^{+7}_{-5}$ & $1.40^{+0.10}_{-0.05}$ & $2.57^{+0.10}_{-0.08}$ \\
12 & $-9.252^{+0.008}_{-0.009}$ & $-8.50^{+0.05}_{-0.06}$ & $0.81^{+0.04}_{-0.08}$ & $105^{+12}_{-23}$ & $1.89^{+0.08}_{-0.05}$ & $3.3^{+0.8}_{-0.5}$ \\
13 & $-9.36^{+0.01}_{-0.01}$ & $-8.34^{+0.1}_{-0.1}$ & $0.92^{+0.03}_{-0.09}$ & $166^{+13}_{-38}$ & $2.02^{+0.09}_{-0.06}$ & $2.8^{+0.8}_{-0.4}$ \\

\hline
    \multicolumn{6}{l}{$N_{{\mathrm H,}\texttt{TBabs}}$ = $5.70_{-0.05}^{+0.10}\times10^{22}$~cm$^{-2}$} \\
\hline
    \multicolumn{6}{l}{$\chi^{2}$ / d.o.f = 1806.3 /1528} \\

\hline\hline
\end{tabular}
\caption {Best fit table for the Gaussian-based model. The fitting results of model \texttt{TBabs} $\times$ (\texttt{pcfabs} $\times$ \texttt{nthComp}$+$ \texttt{gaussians}).  Evolution of parameters with time is plotted in Fig. \ref{fig:par}}
\label{tab:bstfit_gau}
\end{table*}

In Table \ref{tab:bstfit_gau} and \ref{tab:ironline}, we show all the fitting and the neutral and ionized line modeling results of 13 spectra. We also show the evoluation of the parameters in Figure \ref{fig:par}. We used \texttt{cflux} to measure the observed and intrinsic flux in the 1--10\,keV band.

The results show that the interstellar absorption column density, modeled with \texttt{TBabs}, is $N_\mathrm{H} = 5.70 ^{+0.99}_{-0.05}\times 10^{22}\,\mathrm{cm^{-2}}$. Over the same period, the unabsorbed flux increases from $1.15 \times 10^{-9}$ to $3.96 \times 10^{-8}\,\mathrm{erg\,cm^{-2}\,s^{-1}}$ between intervals~1 and~3.
Simultaneously, the column density of the partial-covering absorber (\texttt{pcfabs}) decreases from $N_\mathrm{H} = 1.31^{+0.04}_{-0.15} \times 10^{23}$ to $4.7^{+0.5}_{-0.9} \times 10^{22}\,\mathrm{cm^{-2}}$, while the covering fraction $CvrFract$ drops from $0.953^{+0.003}_{-0.02}$ to $0.45^{+0.03}_{-0.08}$. These trends indicate that the main flare is driven by enhanced intrinsic emission together with a result of reduction in line-of-sight obscuration. The increased luminosity likely clears the obscuring material, allowing a larger fraction of soft photons to escape.

After the main flare, both the observed flux and the intrinsic luminosity decrease significantly. At the same time, the absorption increases dramatically, reaching $N_\mathrm{H} = 2.34^{+0.01}_{-0.03} \times 10^{24}~\mathrm{cm^{-2}}$ with a covering fraction of $CvrFract = 0.95^{+0.02}_{-0.04}$. This heavy obscuration strongly suppresses the observed iron emission lines in interval~4. Interestingly, a short-duration “small flare” occurs between intervals~6 and~9, lasting less than 1000~s. During this event, the observed flux increases by nearly an order of magnitude, from $F \approx 3.24 \times 10^{-10}$ (interval~6) to $F \approx 2.19 \times 10^{-9}\mathrm{erg~cm^{-2}~s^{-1}}$ (interval~8). However, the intrinsic flux remains approximately constant, indicating that the observed flux enhancement is driven by a reduction in local absorption. Consistent with this interpretation, acceptable fits are obtained when all parameters of the \texttt{nthComp} component, except for the normalization, are linked across these intervals, suggesting that the coronal properties remain unchanged during the small flare. Intervals~10--13 also show significant variations in the observed flux, while the unabsorbed flux remains nearly constant, further supporting absorption-driven variability.

\subsection{Iron lines}

A clear evolution of the iron line features is observed across the spectra. As the main flare rises toward its peak, three distinct narrow emission lines near 6.4, 6.7 and 7.0\,keV gradually become prominent. After the peak, as the source flux declines, these lines evolve significantly. In interval~4, no emission lines are detected due to the heavy obscuration, and a strong absorption edge near 7.0\,keV becomes apparent. Intervals~5 and~6 exhibit only two narrow emission lines: one is at 6.4\,keV. However, for interval 5, the other one is at 6.6\,keV and for interval 6 at 7.0\,keV. In intervals~7 and~8, the 6.4\,keV line is no longer detected, while the higher-ionization line at 6.7\,keV begins to reappear from interval~9 onward. In interval~10, three narrow emission lines reappear together with a broad iron line, suggesting a partial recovery of the line-emitting region as the obscuration decreases. In addition to the narrow lines, intervals~1–3 and~9–12 require an additional broad Fe line to adequately fit the residuals in the Fe~K region.

\begin{table*}[ht!]
\centering
\renewcommand{\arraystretch}{1.5}
\begin{tabular}{cccccccc}
\hline\hline  
\shortstack[l]{Interval\\~~}
&
\shortstack[l]{Fe K$\alpha$\\(keV)}
&
\shortstack[l]{EW\\(keV)}&
\shortstack[l]{Fe \textsc{xxv}\\(keV)}
&
\shortstack[l]{EW\\(keV)}
&
\shortstack[l]{Fe \textsc{xxvi} Ly$\alpha$ \\ or Fe K$\beta$ (keV)}
&
\shortstack[l]{EW\\(keV)}
&
\shortstack[l]{Broad Iron Line EW\\  (keV)}\\

\hline

1 & $6.450_{-0.005}$ & $0.05^{+0.01}_{-0.01}$ & - & - & - & - & $0.21^{+0.04}_{-0.04}$\\
2 & $6.450_{-0.005}$ & $0.04^{+0.01}_{-0.01}$  & $6.680^{+0.005}_{-0.008}$ 
& $0.05^{+0.01}_{-0.01}$ & $7.00^{+0.01}_{-0.01}$ & $0.03^{+0.01}_{-0.01}$ & $0.23^{+0.03}_{-0.04}$  \\
3 & $6.45_{-0.01}$ & $0.020^{+0.005}_{-0.005}$ & $6.681^{+0.004}_{-0.005}$ 
& $0.05^{+0.01}_{-0.01}$& $6.99^{+0.01}_{-0.01}$ & $0.03^{+0.01}_{-0.01}$  & $0.18^{+0.04}_{-0.02}$ \\
4 & - & - & - & - & - & - & -\\
5 & $6.39^{+0.05}_{-0.01}$ & $0.09^{+0.02}_{-0.01}$ & $6.60^{+0.03}$ & $0.06^{+0.02}_{-0.01}$ &-& -&- \\
6 & $6.45_{-0.05}$ & $0.21^{+0.05}_{-0.05}$ & - & - & $7.10_{-0.033}$ & $0.09^{+0.07}_{-0.08}$ & -\\
7 & -& - & $6.60^{+0.01}$& $0.13^{+0.04}_{-0.04}$ & - & - & - \\
8 & - & - & - & - & $7.02^{+0.07}_{-0.02}$& $0.13^{+0.01}_{-0.01}$ & - \\
9 & $6.44^{+0.06}_{-0.05}$ & $0.07^{+0.02}_{-0.02}$ & $6.65^{+0.02}_{-0.05}$& 
$0.04^{+0.01}_{-0.01}$  & - & - & $0.78^{+0.15}_{-0.13}$ \\
10 & $6.450_{-0.004}$ & $0.16^{+0.02}_{-0.02}$ & $6.68^{+0.01}_{-0.02}$& 
$0.08^{+0.02}_{-0.02}$   & $7.00^{+0.01}_{-0.01}$ & $0.09^{+0.02}_{-0.02}$ & $0.50^{+0.07}_{-0.08}$ \\
11 & $6.40^{+0.04}_{-0.01}$ & $0.31^{+0.02}_{-0.02}$ & $6.681^{+0.003}_{-0.003}$ 
& $0.21^{+0.02}_{-0.02}$ &$7.000^{+0.004}_{-0.003}$ & $0.21^{+0.02}_{-0.02}$ & $0.40^{+0.06}_{-0.06}$ \\
12 & $6.39^{+0.004}_{-0.004}$ & $0.53^{+0.05}_{-0.05}$ & $6.683^{+0.005}_{-0.007}$ & 
$0.22^{+0.04}_{-0.04}$ & $7.01^{+0.05}_{-0.01}$ & $0.18^{+0.04}_{-0.04}$ & - \\
13 & $6.40^{+0.05}_{-0.01}$  &$0.6^{+0.1}_{-0.3}$ & - & - & $7.00^{+0.01}_{-0.01}$ & $0.15^{+0.05}_{-0.03}$ & - \\

\hline\hline
\end{tabular}
\caption {Neutral and ionized line modeling results. The fitting results of model \texttt{TBabs} $\times$ (\texttt{pcfabs} $\times$ \texttt{nthComp}$+$ \texttt{gaussians}).  The broad iron line energy is fixed at 6.4 keV.}
\label{tab:ironline} 
\end{table*}

%%%%%%%%%%%%%%%%%%%%%%%
\section{Reflection-based Model}\label{sec:ref}

The X-ray spectra of GRS~1915+105 during our observations exhibit strong residuals in the Fe~K band that cannot be fully accounted for by pure Comptonization or absorption models. We test if these features can be fitted using the reflection model \texttt{xillverCp} \cite{Garcia2013}, which describes the reprocessing of the primary continuum by distant material. The residuals include both neutral and highly ionized iron emission lines, which are unlikely to originate from a single reflecting region \cite{Miller_2020}. We thus include two reflection components: a neutral component (\texttt{xillverCp1}) with the ionization parameter fixed at $\log\xi = 0$, and an ionized component (\texttt{xillverCp2}) with $\log\xi$ left free. Both reflection components, together with the primary continuum (\texttt{nthComp}), are partially covered by distinct absorbers modeled with \texttt{pcfabs}. We also tried a multiplicative ionized absorber \texttt{zxipcf} \cite{Millerrr2006,Reeves2008}, but it does not provide a better fit.

\begin{figure}
    \centering
    \includegraphics[width=\linewidth]{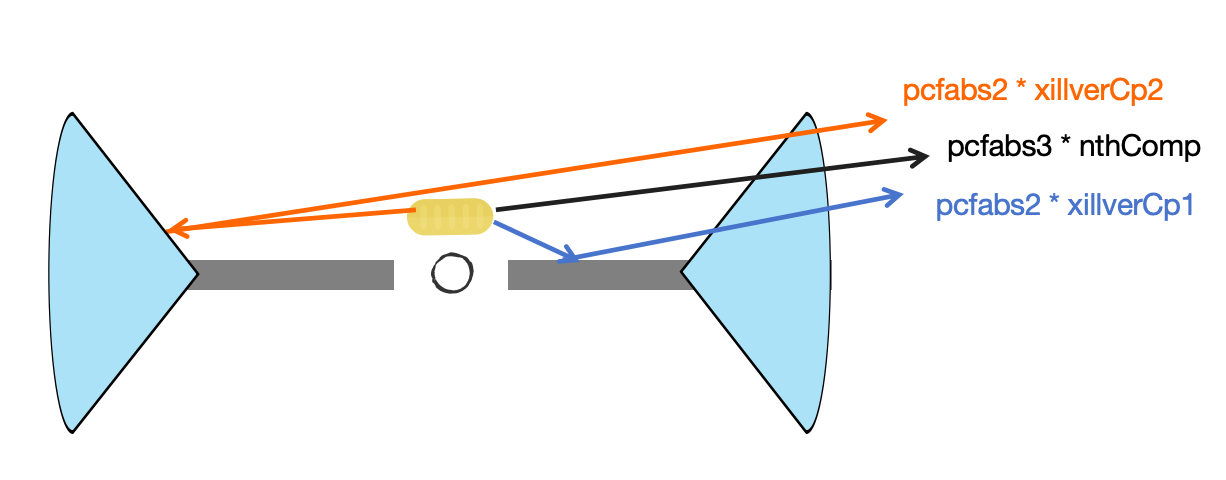}  
    \caption{ The assumed geometry of the reflection-based spectral model. The corona is shown in yellow, and the absorbing material is shown in cyan. The continuum component (\texttt{nthComp}; black) originates in the corona and is absorbed by \texttt{pcfabs3}. The neutral reflection component (\texttt{xillverCp1}; orange) arises from distant, less-ionized material and is absorbed by a separate layer with different properties, modeled with \texttt{pcfabs1}. The ionized reflection component (\texttt{xillverCp2}; blue) originates from highly ionized material and is absorbed by another part of the absorber, \texttt{pcfabs2}.}
    \label{fig:reflect}
\end{figure}

The photon index $\Gamma$ and the electron temperature $\mathrm{kT_ e}$ of both reflection components were linked to those of the primary continuum (\texttt{nthComp}), assuming a common illuminating spectrum. We fixed the inclination angle at 64$^\circ$ \cite{Reid2014}. The iron abundance was tied between the two reflectors and linked across all intervals, while the gas density $\log n$ was allowed to differ between \texttt{xillverCp1} and \texttt{xillverCp2}, reflecting potentially different physical conditions in the two regions. For intervals 10 to 13, we linked \texttt{pcfabs} through different regions to avoid degeneracy. We set \texttt{refl\_frac=-1} in \textsc{xspec} language to model the reflection-only spectra, excluding the direct continuum contribution from the reflection components. All other parameters were treated in the same manner as in the previous modeling. The final spectral model is \texttt{TBabs} $\times$ (\texttt{pcfabs1} $\times$ \texttt{xillverCp1} $+$ \texttt{pcfabs2} $\times$ \texttt{xillverCp2} $+$ \texttt{pcfabs3} $\times$ \texttt{nthComp}) (Reflection-based model). Such geometry (shown in Fig. \ref{fig:reflect}) has also been proposed by \cite{Ratheesh2021} and \cite{Zhou_2025}. To account for the instrumental residuals near the Silicon K-edge, we included a local Gaussian component fixed at $1.84\text{ keV}$ for interval 3 and interval 5. In Appendix~\ref{app-A}, we present the best fit table for this model.

\subsection{Spectral evolution} 

To robustly explore the parameter space and estimate confidence intervals, we employed the Markov Chain Monte Carlo (MCMC) \textsc{xspec} command \texttt{mcmc}, using the Goodman-Weare algorithm. The results show that the trend of the modeled unabsorbed flux is consistent with that of the Gaussian-based model. 

\begin{figure}
    \centering
    \includegraphics[width=\linewidth]{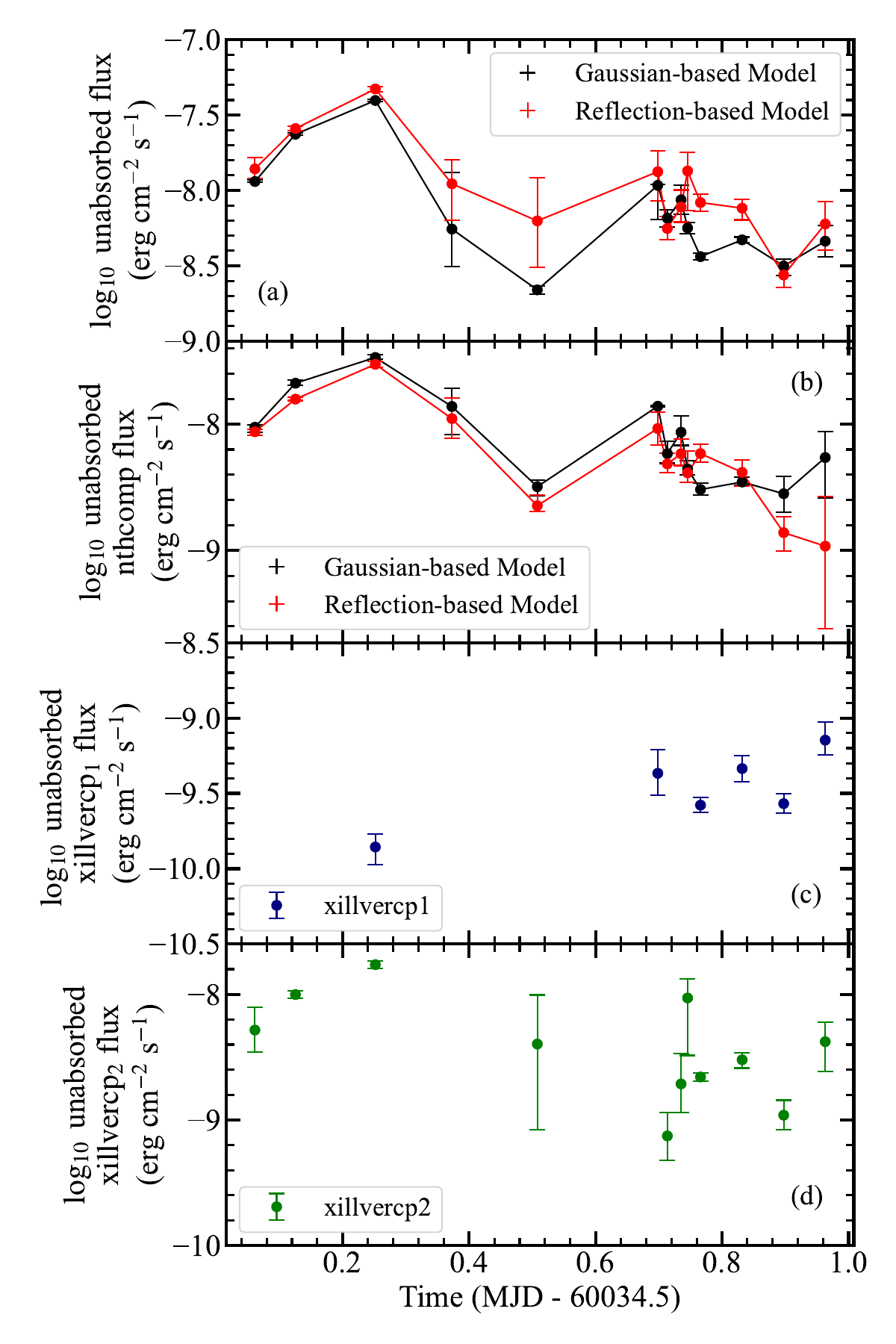}  
    \caption{The evolution of the unabsorbed flux for two different models.The flux are measured by \texttt{cflux} in \textsc{xspec}.  The \texttt{Gaissian}-based model is shown in black, and the reflection-based model is in red.  (a) The total unabsorbed flux. Total flux of reflection-based model is calculated from \texttt{cflux} of the other components. (b) The unabsorbed flux of the continuum component \texttt{nthcomp}. (c) The unabsorbed flux of the neutral reflection component \texttt{xillvercp1} (in blue).  (c) The unabsorbed flux of the inoized reflection component \texttt{xillvercp2} (in green).  }
    \label{fig:xill_gau}
\end{figure}

The results also reveal diverse reflection and absorption structures across the 13 intervals. The iron abundance is measured to be $A_{\mathrm{Fe}} = 4.6^{+0.8}_{-0.3}\,A_{\odot}$. The gas density is constrained to $\log n = 15.2^{+0.8}_{-0.2}$ for the neutral reflector and $\log n = 18.5^{+0.3}_{-0.2}$ for the ionized reflector. These results suggest that the neutral reflector is located farther from the X-ray source than the ionized reflector, as the lower density inferred for \texttt{xillverCp1} is consistent with reflection from more distant, less ionized material.

The evolutionary trend of the continuum flux inferred from the reflection-based model is broadly consistent with that obtained from the Gaussian-based model (see Fig.~\ref{fig:xill_gau}). Interestingly, in some intervals, particularly intervals~13, the reflected flux modeled with \texttt{xillverCp} exceeds the continuum flux. This behavior likely reflects the strong contribution from reflection as well as differences between the phenomenological Gaussian approach and the self-consistent reflection modeling. Overall, these results confirm that the observed flux variability is mainly driven by changes in the intrinsic luminosity for the main flare and changes in the local absorption for the later decline.

\section{Discussion}\label{sec:dis}

Flare events are frequently observed during the obscured state of GRS~1915+105, and several interpretations have been proposed. Time-resolved NICER spectroscopy showed that rapid changes in column density and ionization within a stratified absorber can reproduce apparent X-ray brightening, while Chandra observations support a failed-wind scenario in which dense wind material intermittently obscures or reveals the central source \citep[e.g.,][]{Neilsen2020,Miller_2020}. Short flares may therefore be associated with changes in absorber geometry, whereas longer re-brightenings could also involve enhanced mass inflow or disk instabilities~\citep{Athulya2023}. The detection of strong radio activity further indicates that accretion--jet coupling remains active during the obscured state, and some flares may be related to jet activity or to a vertically extended corona temporarily clearing the obscurer~\citep{motta2021}. Radio observations also revealed a large misalignment between the continuous jet and discrete ejecta. The flares are interpreted as discrete ejection events launched from its distorted disk~\citep{jiang2026}. In addition, AstroSat/NuSTAR observations have shown that some flare spectra can be described either by highly ionized absorption or by relativistically blurred reflection, highlighting the degeneracy between absorption, reflection, and intrinsic continuum variations~\citep{Boked2024}. These studies suggest that flares in the obscured state may involve several coupled processes, including variable absorption, intrinsic luminosity changes, reflection, and jet-related activity. However, how the continuum, local absorber, and reflection components evolve together during an individual bright flare remains unclear. Our work addresses this point by following the spectral evolution of the 2023 flare with time-resolved NICER spectroscopy.

An alternative explanation for rapid spectral variability is occultation by local dense clumps crossing the line of sight. Such clumpy absorbers have been invoked in several BHXRBs. For example, during the 2015 outburst of V404~Cyg, plateau and low-flux states showed Compton-thick partial covering and strong reflection features, which can be explained by a geometrically thick inner flow launching a clumpy outflow \citep{Motta2017,Kajava2020}. In Cyg~X-3, the Wolf--Rayet companion provides a dense and clumpy wind that produces strong absorption and iron-line features across different states \citep{van1996,Koljonen2020}. These analogies show that clumpy obscuration is a plausible alternative, although in GRS~1915+105 the correlated evolution of absorption, continuum, and reflection during the 2023 flare favors a more coherent, stratified wind geometry.

\subsection{Flux--Absorption Interplay}

We performed a time-resolved spectral analysis of GRS~1915+105 during its obscured state using NICER and Swift observations of an X-ray flare between 2023 March~31 and April~1. Our results reveal a complex interplay between the intrinsic emission, reflection, and the local absorber.

\begin{figure}
    \centering
    \includegraphics[width=\linewidth]{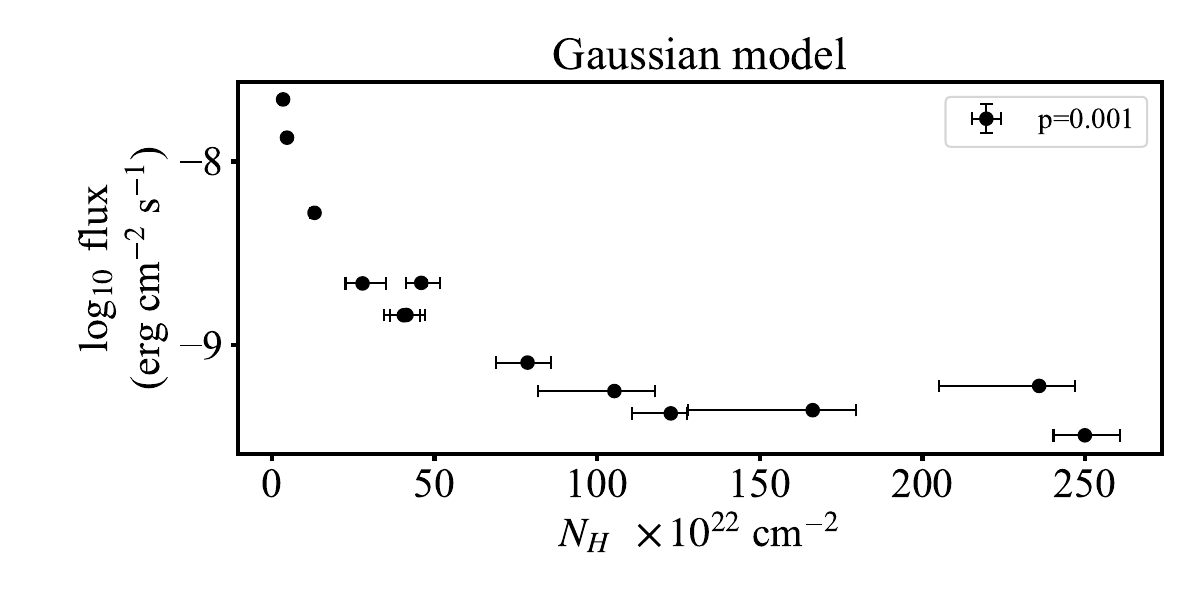} 
    \includegraphics[width=\linewidth]{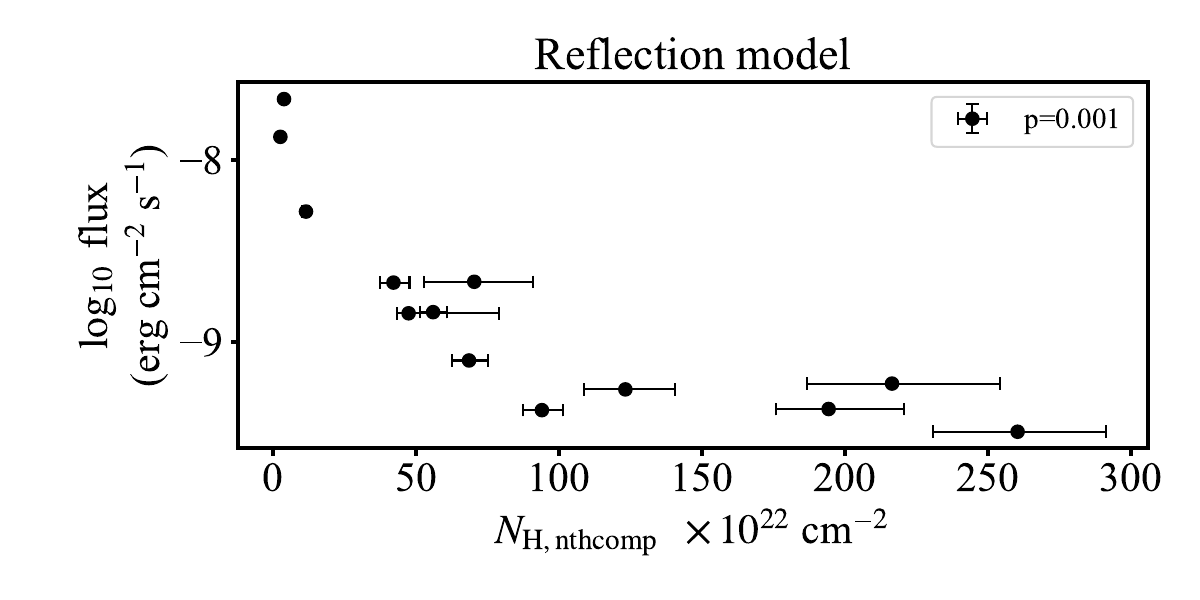}
    \caption{Pearson correlation coefficient test for the relation between the observed X-ray flux and column density of the local absorber $N_\mathrm{H}$. Upper panel: The relation between $N_\mathrm{H}$ and observed flux for the Gaussian-based model. Bottom panel: The relation between $N_\mathrm{H,3}$ for \texttt{nthcomp} and observed flux for reflection-based model.} 
    
    \label{fig:xill_pea}
\end{figure}

To get the relation between the observed X-ray flux and the absorber properties, we performed a Pearson correlation coefficient test, shown in Fig.~\ref{fig:xill_pea}. The Gaussian-based model show a statistically significant  ($>3\sigma$)  anti-correlation between the column density $N_{\mathrm {H}}$ and the observed flux. Moreover, the reflection-based modeling shows a similar result. The observed flux also shows a strong anti-correlation with the column density $N_{\mathrm {H,nthcomp}}$ of the absorber that acts on the continuum component. This suggests that variations in the continuum absorber play a dominant role in regulating the observed flux, while the absorbers associated with the reflection components have a comparatively smaller impact. This behavior indicates that the observed X-ray variability is driven not only by intrinsic changes in the accretion flow, but also by rapid variations in line-of-sight obscuration. One physical interpretation is that the enhanced radiation from the central source may partially disperse or ionize the obscuring material, thereby reducing its effective covering fraction along the line of sight. Similar anti-correlations between absorption and flux have been reported in previous studies of the obscured state (e.g., \cite{Neilsen2020,Athulya2023}), supporting a significant role of variable obscuration in shaping the X-ray phenomenology of GRS~1915+105.

The spectra are strongly affected by local absorption and reflection, so degeneracies between the continuum shape, absorber parameters, and reflection strength are unavoidable. The column density and covering fraction of the absorber can partially mimic changes in the continuum slope and reflection. We therefore do not interpret individual best-fit parameters in isolation. Instead, our conclusions are based on the joint temporal evolution of total 13 intervals, together with the persistence of Fe-K residuals under different model configurations.

To further examine parameter degeneracies, we showed MCMC results for four representative spectra selected from different phases of the flare: the main flare (Interval 2), the Compton-thick obscured phase (Interval 5), the small flare (Interval 8), and the late reflection-dominated phase (Interval 10). The results are shown in Appendix~\ref{mcmc}. The MCMC posteriors show the expected covariance between absorber column density, covering fraction, $\Gamma$, $\log \xi$, and other parameters. We interpreted the flux-absorption interplay as a physically motivated explanation that consistently accounts for the observed temporal evolution, rather than as a unique solution.

\subsection{Iron Emission Line}

The Fe~K emission lines also show significant evolution during the observations. One possible explanation for this variability is absorption by local, dense clumps crossing the line of sight. The stochastic motion of such clumps can lead to time-dependent changes in the strength and visibility of the iron-line features. Similar clumpy obscuration scenarios have been proposed for several BHXRBs. During its 2015 outburst, V404~Cyg displayed plateau and low-flux states characterized by Compton-thick partial covering and strong reflection features. These spectra have been interpreted in terms of a geometrically thick inner flow launching a clumpy outflow (e.g., \cite{Motta2017,Kajava2020}). Similarly, Cyg~X-3, which hosts a Wolf–Rayet donor star \cite{van1996}, exhibits strong absorption and prominent iron features across all states, likely associated with clumpy material in the stellar wind \cite{Koljonen2020}.

However, no significant correlations are found between the absorber properties, the emission line flux, and the disk ionization parameter $\log \xi$ in either the Gaussian-based or reflection-based models. This indicates that the iron-line variability is not governed by a single controlling parameter, but instead supports a geometry-dominated reflection scenario, in which multiple reflection regions with distinct absorbers become visible or obscured as the source–absorber configuration evolves.

Furthermore, the reflection-based model shows that the neutral and ionized reflection components are not simultaneously visible and are subject to different absorbers. Strong reflection features do not always accompany a high intrinsic continuum flux. This suggests that the visibility of reflection is primarily determined by whether the reflecting regions are obscured or exposed along the line of sight, rather than by changes in the intrinsic coronal emission.

Using the measured $\log \xi$ and luminosity, and using the disk density of $ n_{e}$ from the fitting results of both neutral and ionized reflection components, we can roughly estimate the distance of the reflector as $R = \sqrt{L / (n\xi)}$ \cite{Tarter1969}.  For the ionized reflector, $n_{e} \sim 10^{18}~\mathrm{cm^{-3}}$, the inferred $R_{\mathrm{xillvercp2}}$ is $ \sim 10^{7}$~cm. This comparison suggests that the heavily obscured state may be due to the inner failed wind discussed by \cite{Miller_2020}.

\subsection{Evolution of the Obscuration}

Our results extend previous studies of the obscured state by resolving the spectral evolution of one bright flare into a sequence of physically connected stages. Within the framework of the failed disk wind scenario proposed by \citet{Miller_2020}, the flare we observed can be interpreted by a later stage of the same long-lived episode, in which the obscuring material partially thins or reconfigures, intermittently revealing distinct reflection regions. The observed spectral evolution can be interpreted in terms of transitions between three characteristic geometric states of the obscurer.

Phase I: Brightening with reduced obscuration.
At the onset of the flare, enhanced coronal emission likely reduces the effective obscuration, allowing both ionized and neutral reflection regions to become visible through the residual failed-wind material.

Phase II: Fading under deep obscuration.
As the intrinsic luminosity decreases, the radiation pressure weakens and can no longer effectively counteract gravity, leading to a re-thickening of the obscuring material along the line of sight. The obscuration strengthens and becomes Compton-thick, strongly suppressing the observed continuum and rendering most iron-line features undetectable.

Phase III: Reflection-dominated visibility.  
At later times, the intrinsic luminosity remains low and the obscuration is still strong, but changes in the absorber geometry allow reflection regions to become visible while the direct continuum remains suppressed. This phase reflects a geometry-driven reconfiguration of the obscurer rather than a simple decrease in column density.

\subsection{Radio flares}

Multi-wavelength monitoring of GRS~1915+105 reveals a close temporal connection between X-ray flares and radio activity, indicating a coupling between the accretion flow and jet ejection. During the obscured state in 2023, a short X-ray flare detected by \textit{Swift}/BAT on March~25 (MJD~60028.5) rapidly rose to a 15--50~keV flux of $\sim1.6~\mathrm{counts~cm^{-2}~s^{-1}}$ within $\sim9$~hours and decayed on a timescale of $\sim3$~hours \cite{Trushkin_2023_2}. A radio flare detected by RATAN-600 on March~27 (MJD~60030.18) followed this event with a delay of $\sim1.6$~days. The larger X-ray flare analyzed in this work occurred on March~29 (MJD~60034.67) and was followed by a major radio outburst exceeding 2~Jy at 21.7~GHz on April~3 (MJD~60037.16; \cite{Trushkin_2023_3}), corresponding to a delay of $\sim2.5$~days. The recurrence of radio flares following X-ray re-brightenings, with comparable time lags, suggests that these events are not coincidental but instead reflect a physical connection between X-ray activity and jet launching.

\begin{figure}
    \centering
    \includegraphics[width=1\linewidth]{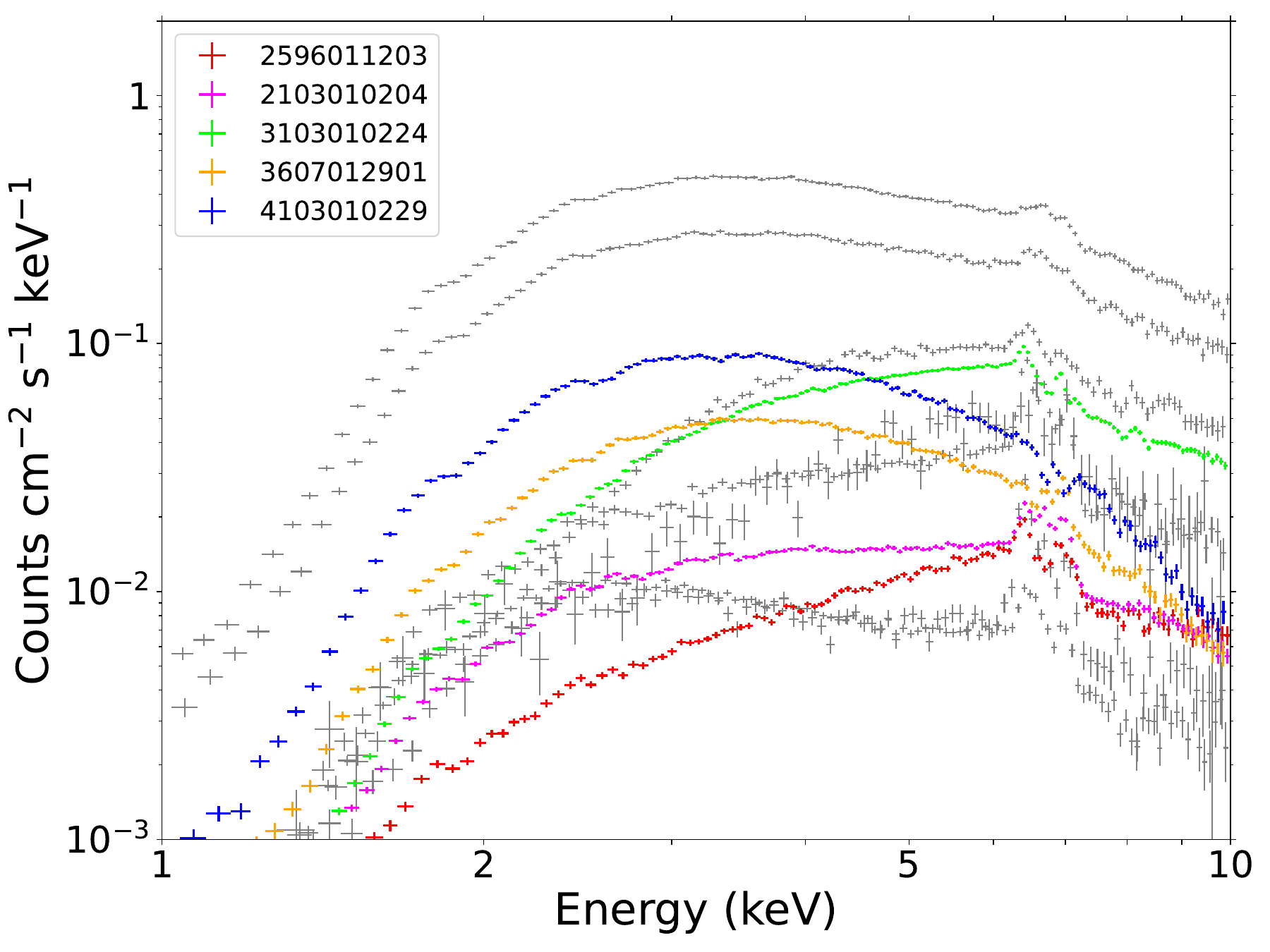} 
    \caption{Comparison of five previously reported flares from \citet{Athulya2023} with five representative spectra. The spectra of our study are shown in gray. The spectrum from flare peak of the re-brightening phases  are ObsID 2596011203 (red, long re-brightening), ObsID 2103010204 (magenta, short re-brightening), ObsID 3103010224 (lawn-green, short re-brightening), ObsID 3607012901 (orange, long re-brightening), and ObsID 4103010229 (blue, long re-brightening). The order of ObsID is based on time of re-brightening in Fig.~\ref{fig:maxi}.}
    \label{fig:other}
\end{figure}

Such X-ray--radio associations are commonly observed during the obscured state. Figure~\ref{fig:other} compares our observations with several previously reported flare events accompanied by radio activity from \citet{Athulya2023}. These events span a wide range of timescales, from short re-brightenings lasting a few thousand seconds to long-duration flares persisting for several days, and exhibit diverse spectral shapes and peak fluxes. The radio emission therefore provides an important diagnostic of jet activity and outflow dynamics. Long-duration X-ray flares are likely associated with mini-outbursts driven by thermal--viscous instabilities in the outer accretion disk, while shorter, rapid X-ray flares might be linked to transient jet ejections that temporarily disrupt or clear the obscuring material \cite{Athulya2023}.

Within this framework, X-ray flares trace changes in the inner accretion flow and/or obscuration geometry, while the delayed radio flares reflect the subsequent propagation of relativistic ejecta, providing direct evidence for accretion--jet coupling during the obscured state.

\section{Conclusions}\label{sec:conclusion}

We have analyzed NICER and Swift observations of a large X-ray flare from GRS~1915+105 during its obscured state. The data were divided into 13 time intervals and modeled using time-resolved spectroscopy, employing both phenomenological Gaussian-based and physically motivated reflection-based models. Our results show that the continuum and reflection components are subject to distinct absorbers, and that their visibility evolves on timescales of hundreds of seconds. During the main flare, the observed brightening is driven by intrinsic luminosity increase, accompanied by changes in absorber geometry that further enhance the observed flux and give rise to prominent iron emission lines. As the flare evolves, the dominant driver shifts: the subsequent flux decay is largely governed by increased obscuration, which suppresses the continuum and alters the visibility of the reflection components. These observations are consistent with a post–failed-wind re-illumination phase, in which the central source becomes temporarily visible through a Compton-thick, clumpy obscurer. 

We also interpret the observed behavior in terms of a three-phase evolution of the obscuring material: an initial brightening phase in which enhanced coronal emission temporarily reduces the effective obscuration and reveals multiple reflection regions; a subsequent fading phase characterized by re-thickening of the line-of-sight absorber and strong suppression of both the continuum and iron-line emission; and a later reflection-dominated phase, in which changes in absorber geometry allow reflected emission to remain visible while the direct continuum stays obscured.

Future multi-wavelength campaigns with higher spectral resolution will be essential to test the accretion–jet coupling scenario more directly and to further constrain the geometry of the obscuring material. Beyond GRS~1915+105, this framework provides a useful avenue for interpreting obscured accretion states in other X-ray binaries, and may offer broader insights into the interplay between winds, jets, and obscuration in both black hole binaries and obscured active galactic nuclei.

\begin{acknowledgments}
      This work was supported by the National Natural Science Foundation of China (NSFC), Grant No.~W2531002.
      SZ also acknowledges the support from the China Scholarship Council (CSC), Grant No.~202406100028.
\end{acknowledgments}

\section*{Data Availability}

The data that support the findings of this article are openly available~\footnote{\href{https://heasarc.gsfc.nasa.gov/docs/archive.html}{https://heasarc.gsfc.nasa.gov/docs/archive.html}}.

\appendix

\section{Best fit table for the Reflection-based model} \label{app-A}

The best fit values for the reflection-based model are reported in reported in Tables~\ref{tab:bstfit_ref1} and \ref{tab:bstfit_ref2}.

\begin{table*}[ht!]
\caption{Best fit table for the Reflection-based model (part 1)}
\label{tab:bstfit_ref1} 
\centering
\renewcommand{\arraystretch}{1.5}
\begin{tabular}{ccccccccccccccccccc}
\hline\hline  
\shortstack[l]{Interval \\~~}
&
\shortstack[l]{$\log_{10}F$\\($\mathrm{erg\,cm^{-2}\,s^{-1}}$)}
&
\shortstack[l]{$\log_{10}F_{\mathrm{unabs}}$\\($\mathrm{erg\,cm^{-2}\,s^{-1}}$)}
&
\shortstack[l]{$\log_{10}F_{\mathrm{unabs,xillvercp1}}$\\($\mathrm{erg\,cm^{-2}\,s^{-1}}$)}
&
\shortstack[l]{$\log_{10}F_{\mathrm{unabs,xillvercp2}}$\\($\mathrm{erg\,cm^{-2}\,s^{-1}}$)}
&
\shortstack[l]{$\log_{10}F_{\mathrm{unabs,nthcomp}}$\\($\mathrm{erg\,cm^{-2}\,s^{-1}}$)}
&
\shortstack[l]{$CvrFract_\mathrm{xillvercp1}$ \\~~}
 \\

\hline

Interval 1 & $-8.280^{+0.003}_{-0.003}$ & $-7.85^{+0.08}_{-0.06}$ & - & $-8.3^{+0.2}_{-0.2}$ & $-8.06^{+0.02}_{-0.03}$ & -\\
Interval 2 & $-7.869^{+0.001}_{-0.001}$ & $-7.59^{+0.01}_{-0.01}$ & - & $-8.00^{+0.03}_{-0.03}$ & $-7.80^{+0.01}_{-0.01}$ & -\\
Interval 3 & $-7.661^{+0.001}_{-0.001}$ & $-7.32^{+0.02}_{-0.02}$ & $-9.86^{+0.09}_{-0.12}$ & $-7.76^{+0.03}_{-0.03}$ & $-7.53^{+0.02}_{-0.02}$ & - \\
Interval 4 & $-9.21^{+0.01}_{-0.01}$ & $-8.0^{+0.2}_{-0.2}$ & - & - & $-8.0^{+0.2}_{-0.2}$ & -\\
Interval 5 & $-9.373^{+0.003}_{-0.003}$ & $-8.2^{+0.3}_{-0.3}$ & - & $-8.4^{+0.4}_{-0.7}$ & $-8.64^{+0.09}_{-0.04}$ & -\\
Interval 6 & $-9.48^{+0.01}_{-0.01}$ & $-7.9^{+0.1}_{-0.1}$ & $-10.33^{+0.07}_{-0.08}$ & - & $-7.9^{+0.1}_{-0.1}$ & -\\
Interval 7 & $-9.09^{+0.01}_{-0.01}$ & $-8.2^{+0.08}_{-0.08}$ & - & $-9.1^{+0.2}_{-0.2}$ & $-8.31^{+0.09}_{-0.07}$ & -\\
Interval 8 & $-8.65^{+0.01}_{-0.01}$ & $-8.1^{+0.1}_{-0.1}$ & - & $-8.7^{+0.2}_{-0.2}$ & $-8.23^{+0.11}_{-0.09}$ & -\\
Interval 9 & $-8.838^{+0.005}_{-0.005}$ & $-7.9^{+0.1}_{-0.3}$ & - & $-8.0^{+0.2}_{-0.5}$ & $-8.39^{+0.17}_{-0.07}$ & -\\
Interval 10 & $-8.833^{+0.003}_{-0.003}$ & $-8.08^{+0.05}_{-0.06}$ & $-9.58^{+0.05}_{-0.05}$ & $-8.66^{+0.03}_{-0.03}$ & $-8.23^{+0.07}_{-0.07}$& =$CvrFract_\mathrm{nthcomp}$ \\
Interval 11 & $-8.666^{+0.003}_{-0.003}$ & $-8.12^{+0.06}_{-0.08}$ & $-9.34^{+0.08}_{-0.09}$ & $-8.52^{+0.05}_{-0.07}$ & $-8.4^{+0.1}_{-0.1}$& =$CvrFract_\mathrm{nthcomp}$ \\
Interval 12 & $-9.256^{+0.006}_{-0.006}$ & $-8.6^{+0.1}_{-0.1}$ & $-9.57^{+0.07}_{-0.06}$ & $-9.0^{+0.1}_{-0.1}$ & $-8.9^{+0.1}_{-0.1}$ & =$CvrFract_\mathrm{nthcomp}$\\
Interval 13 & $-9.361^{+0.008}_{-0.009}$ & $-8.2^{+0.1}_{-0.1}$ & $-9.1^{+0.1}_{-0.1}$ & $-8.4^{+0.2}_{-0.2}$ & $-9.22^{+0.02}_{-0.02}$& =$CvrFract_\mathrm{nthcomp}$ \\

\hline\hline

\end{tabular}
\end{table*}

\begin{table*}[ht!]
\caption{Best fit table for the Reflection-based model (part 2)}
\label{tab:bstfit_ref2} 
\centering
\renewcommand{\arraystretch}{1.4}
\begin{tabular}{ccccccccccccccccccc}
\hline  \hline 
\shortstack[l]{Interval \\~~}
&
\shortstack[l]{$CvrFract_\mathrm{xillvercp2}$ \\~~}
&
\shortstack[l]{$CvrFract_\mathrm{nthcomp}$ \\~~}
&
\shortstack[l]{$N_{\mathrm {H,xillvercp1}}$\\($10^{22}\,\mathrm{cm^{-2}}$)}
&
\shortstack[l]{$N_{\mathrm {H,xillvercp2}}$\\($10^{22}\,\mathrm{cm^{-2}}$)}
&
\shortstack[l]{$N_{\mathrm {H,nthcomp}}$\\($10^{22}\,\mathrm{cm^{-2}}$)}
&
\shortstack[l]{$\log \xi$\\~~}
&
\shortstack[l]{$\Gamma$ \\~~}
&
\shortstack[l]{$\mathrm{kT_ e}$\\($\mathrm{keV}$)} \\

\hline 

Interval 1 & $0.99^{+0.01}_{-0.01}$ & $0.95^{+0.01}_{-0.01}$ & - & $106^{+45}_{-56}$ & $11.4^{+0.9}_{-1.2}$ & $1.7^{+0.2}_{-0.2}$ & $1.32^{+0.04}_{-0.05}$ & $2.6^{+0.1}_{-0.2}$ \\
Interval 2 & $0.38^{+0.06}_{-0.06}$ & $0.94^{+0.04}_{-0.09}$ & - & $40^{+16}_{-13}$ & $2.5^{+0.3}_{-0.2}$ & $3.16^{+0.01}_{-0.01}$ & $1.46^{+0.02}_{-0.01}$ & $319^{+58}_{-85}$ \\
Interval 3 & $0.58^{+0.09}_{-0.10}$ & $0.79^{+0.09}_{-0.07}$ & - & $52^{+9}_{-8}$ & $3.8^{+0.4}_{-0.4}$ & $3.2^{+0.02}_{-0.01}$ & $1.63^{+0.03}_{-0.03}$ & $400$ \\
Interval 4 & - & $0.93^{+0.02}_{-0.03}$ & - & - & $216^{+38}_{-30}$ & - & $1.75^{+0.05}_{-0.06}$ & $1.7^{+0.3}_{-0.2}$ \\
Interval 5 & $0.998^{+0.001}_{-0.002}$ & $0.74^{+0.02}_{-0.02}$ & - & $308^{+70}_{-62}$ & $94^{+7}_{-6}$ & $1.85^{+0.09}_{-0.11}$ & $2.00^{+0.03}_{-0.03}$ & $47^{+222}_{-44}$ \\
Interval 6 & - & $0.96^{+0.01}_{-0.01}$ & - & - & $260^{+31}_{-30}$ & - & $2.50^{+0.05}_{-0.05}$ & $2.2^{+0.2}_{-0.1}$ \\
Interval 7 & $0.92^{+0.06}_{-0.11}$ & $0.90^{+0.02}_{-0.02}$ & - & $34^{+45}_{-20}$ & $68^{+7}_{-6}$ & $2.3^{+0.1}_{-0.1}$ & $2.2^{+0.1}_{-0.1}$ & $134^{+181}_{-131}$ \\
Interval 8 & =$CvrFract_\mathrm{nthcomp}$ & $0.90^{+0.03}_{-0.03}$ & - & =$N_\mathrm{H,nthcomp}$  & $42^{+6}_{-5}$ & $2.0^{+0.4}_{-0.5}$ & $1.4^{+0.3}_{-0.1}$ & $1.9^{+0.4}_{-0.2}$ \\
Interval 9 & $0.99^{+0.01}_{-0.02}$ & $0.85^{+0.05}_{-0.03}$ & - & $136^{+34}_{-99}$ & $47^{+32}_{-4}$ & $2.16^{+0.08}_{-0.09}$ & $1.54^{+0.10}_{-0.07}$ & $2.2^{+0.1}_{-0.1}$ \\
Interval 10 & $0.81^{+0.04}_{-0.03}$ & $0.99^{+0.01}_{-0.02}$ & =$N_\mathrm{H,nthcomp}$ & $6.4^{+0.9}_{-0.7}$ & $56^{+5}_{-4}$ & $3.17^{+0.06}_{-0.06}$ & $2.30^{+0.10}_{-0.08}$ & $168^{+157}_{-125}$ \\
Interval 11 & $0.97^{+0.02}_{-0.07}$ & $0.81^{+0.05}_{-0.06}$ & =$N_\mathrm{H,nthcomp}$ & $16^{+2}_{-3}$ & $70^{+20}_{-17}$ & $3.16^{+0.03}_{-0.03}$ & $1.78^{+0.08}_{-0.07}$ & $29^{+114}_{-15}$ \\
Interval 12 & =$CvrFract_\mathrm{nthcomp}$ & =$CvrFract_\mathrm{nthcomp}$ & =$N_\mathrm{H,nthcomp}$ & =$N_\mathrm{H,nthcomp}$ & $123^{+17}_{-14}$ & $3.10^{+0.08}_{-0.14}$ & $1.84^{+0.06}_{-0.09}$ & $161^{+161}_{-114}$ \\
Interval 13 & =$CvrFract_\mathrm{nthcomp}$ & =$CvrFract_\mathrm{nthcomp}$ & =$N_\mathrm{H,nthcomp}$ & =$N_\mathrm{H,nthcomp}$ & $194^{+26}_{-18}$ & $3.8^{+0.2}_{-0.2}$ & $2.10^{+0.08}_{-0.1}$ & $186^{+144}_{-132}$ \\

\hline
    \multicolumn{9}{l}{$N_{{\mathrm H,}\texttt{TBabs}}$ = $5.96_{-0.06}^{+0.05} \times 10^{22}~\texttt{cm}^{-2}$} \\

    \multicolumn{9}{l}{$A_{{\mathrm Fe}}$ = $4.6_{-0.3}^{+0.8}$~$A_{{\odot}}$} \\

    \multicolumn{9}{l}{$\log n_{neutral} = 15.2^{+0.8}_{-0.2}$} \\

    \multicolumn{9}{l}{$\log n_{ionized} = 18.5^{+0.3}_{-0.2}$} \\
    
\hline
    \multicolumn{6}{l}{$\chi^{2}$ / d.o.f = 1788.25 / 1548} \\ 
\hline

\end{tabular}
\end{table*}

\section{Markov Chain Monte Carlo}\label{mcmc}

\begin{figure*}
    \centering
    \includegraphics[width=0.95\linewidth]{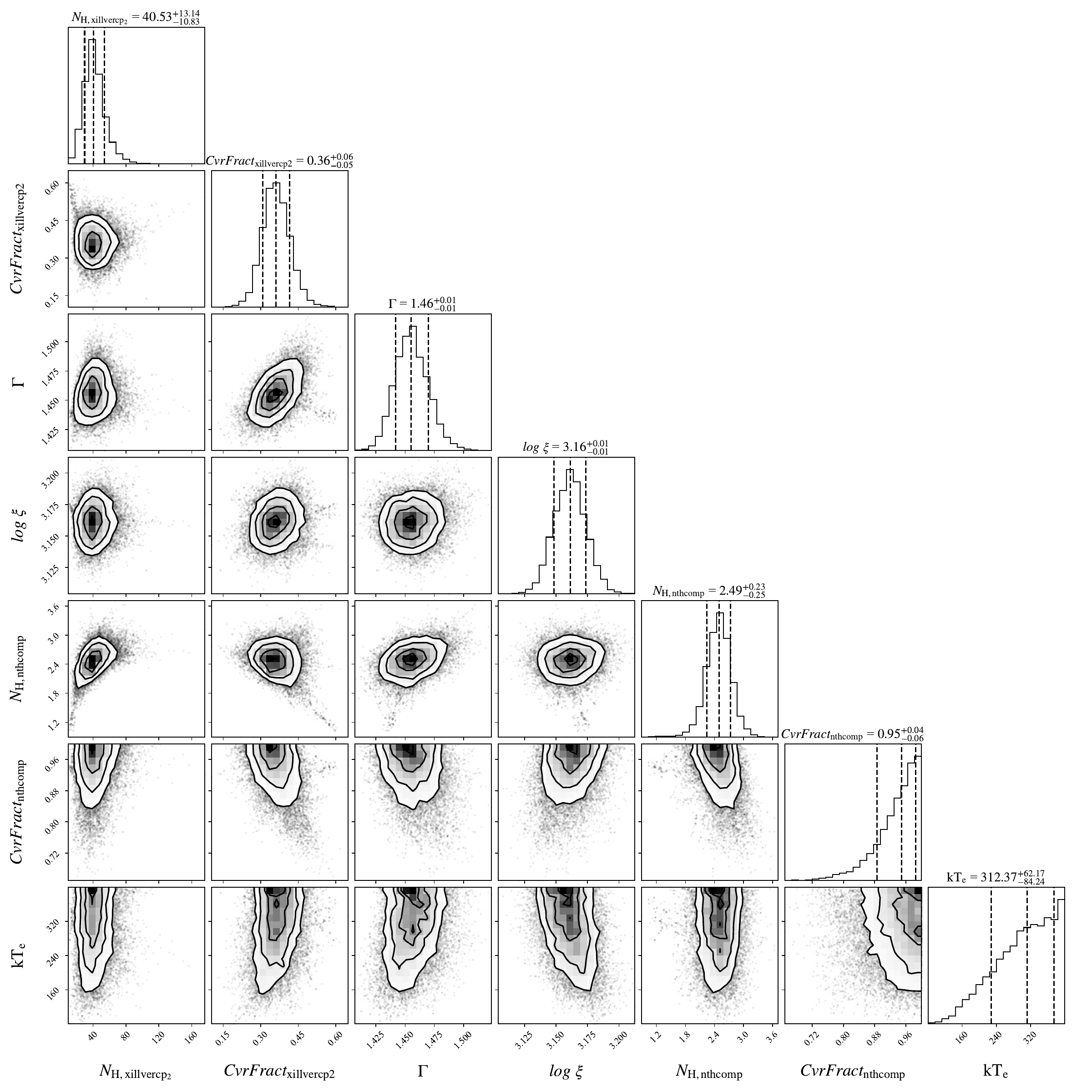} 
    \caption{Corner plot for the free parameter pairs for Interval~2 after the MCMC run.
    \label{fig:mcmc_2}}
\end{figure*}

\begin{figure*}
    \centering
    \includegraphics[width=0.95\linewidth]{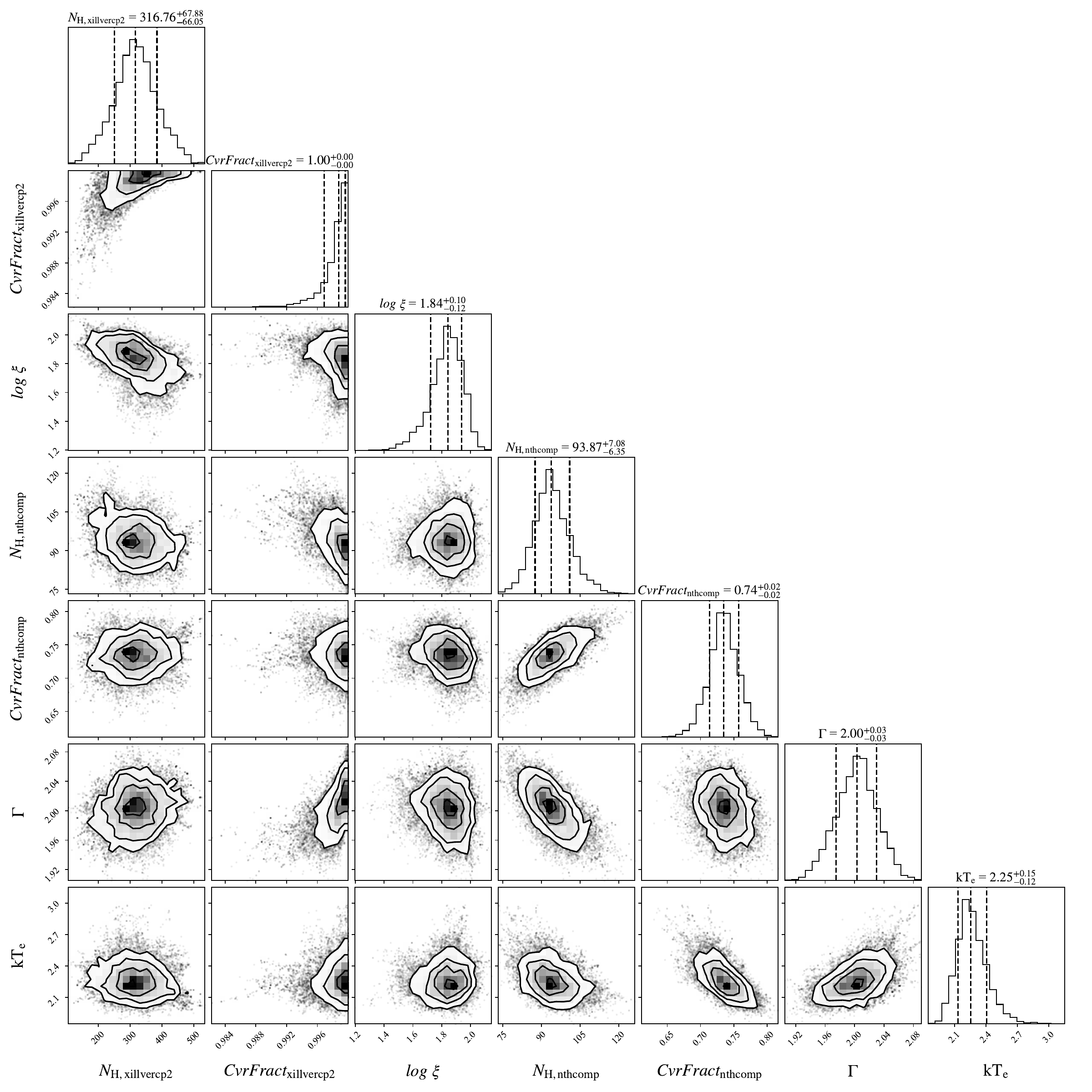} 
    \caption{Corner plot for the free parameter pairs for Interval~5 after the MCMC run.
    \label{fig:mcmc_5}}
\end{figure*}

\begin{figure*}
    \centering
    \includegraphics[width=0.95\linewidth]{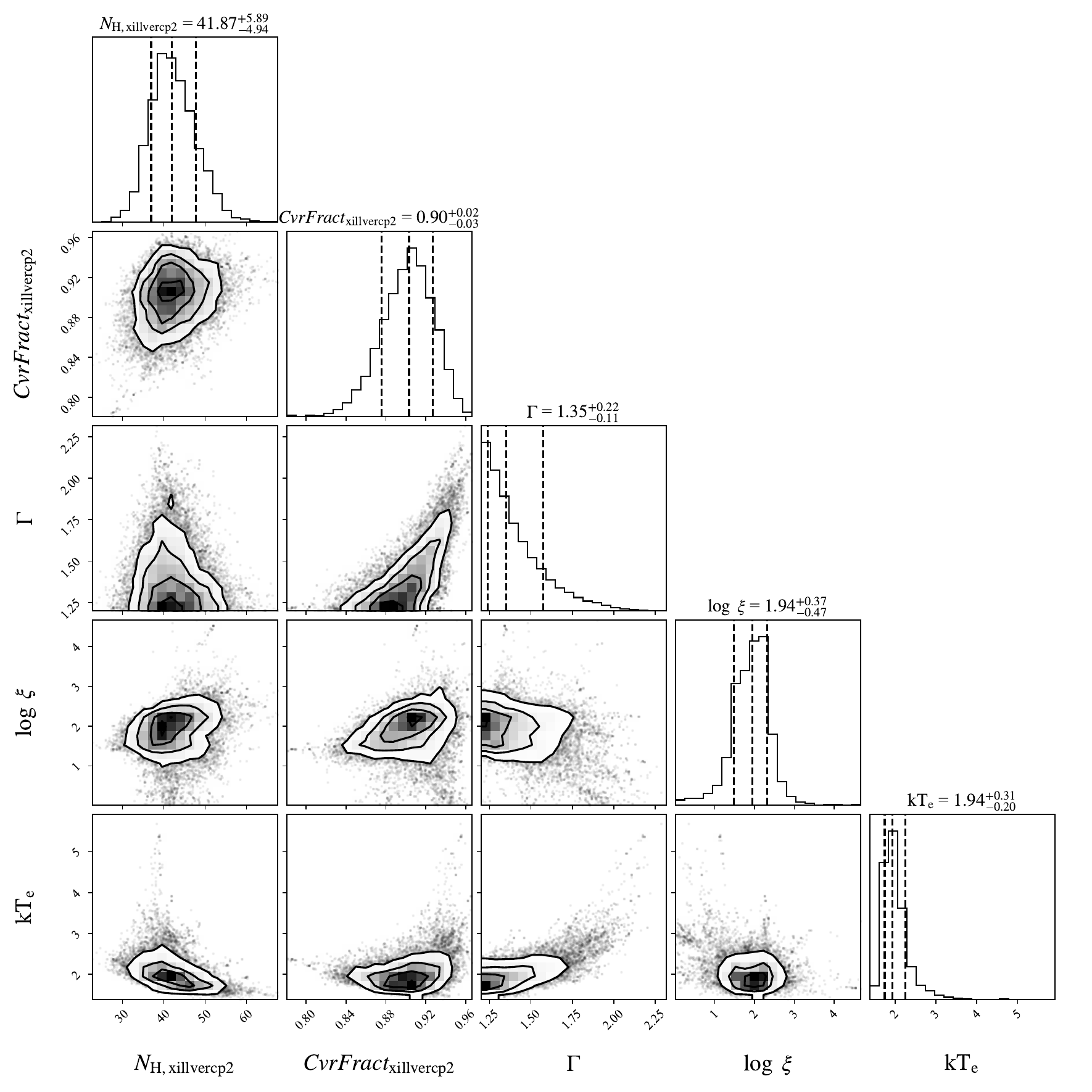} 
    \caption{Corner plot for the free parameter pairs for Interval~8 after the MCMC run.
    \label{fig:mcmc_8}}
\end{figure*}

\begin{figure*}
    \centering
    \includegraphics[width=0.95\linewidth]{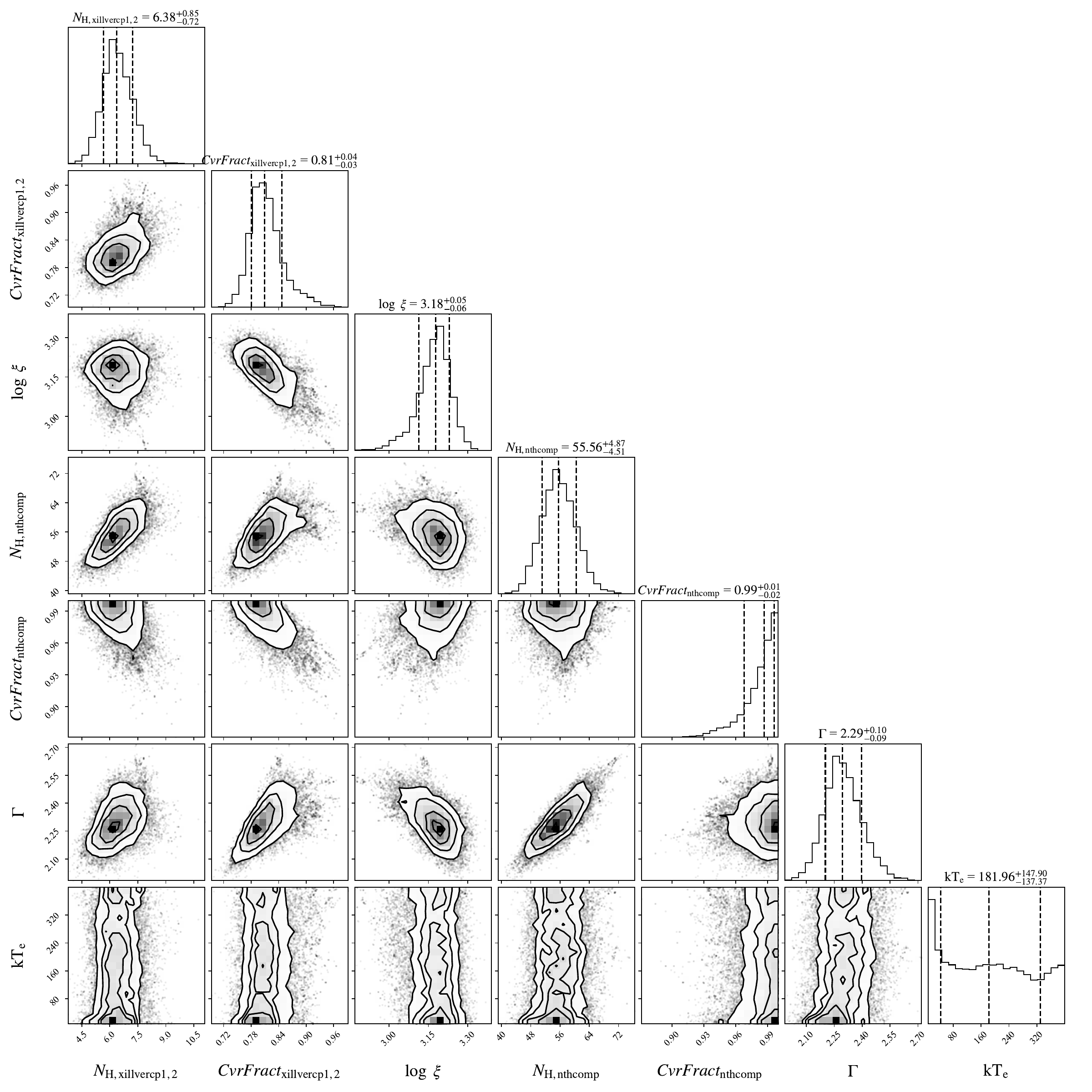} 
    \caption{Corner plot for the free parameter pairs for Interval~10 after the MCMC run.
    \label{fig:mcmc_10}}
\end{figure*}

We employed the Markov Chain Monte Carlo in \textsc{xspec} to analyze four representative spectra selected from different phases of the flare: the main flare (Interval 2), the Compton-thick obscured phase (Interval 5), the small flare (Interval 8) and Interval 10. Figures~\ref{fig:mcmc_2}-\ref{fig:mcmc_10} show the results of our MCMC analysis.

\bibliography{apssamp}% Produces the bibliography via BibTeX.

\end{document}